\documentclass[aps,pra, twocolumn]{revtex4}
\usepackage{graphicx}
\usepackage{bm}
\usepackage{amsmath}
\usepackage[colorlinks=true,allcolors=blue]{hyperref}

\providecommand{\be}{\begin{equation}}
\providecommand{\ee}{\end{equation}}
\providecommand{\ba}{\begin{eqnarray}}
\providecommand{\ea}{\end{eqnarray}}

\usepackage{mathbbol}

\begin{document}

\title{Connecting measurement invasiveness to optimal metrological scenarios}

\author{ Saulo V. Moreira $^1$, Gerardo Adesso$^2$, Luis A. Correa$^2$, Thomas Coudreau$^1$,  Arne Keller$^3$ and P\'erola Milman$^1$}

\affiliation{$^{1}$Univ. Paris Diderot, Sorbonne Paris Cit\'e, Laboratoire Mat\'eriaux et Ph\'enom\`enes Quantiques, UMR 7162, CNRS, F-75205, Paris, France}
\affiliation{$^{2}$Centre for the Mathematics and Theoretical Physics of Quantum Non-Equilibrium Systems (CQNE), School of Mathematical Sciences, The University of Nottingham, University Park, Nottingham NG7 2RD, United Kingdom}
\affiliation{$^{3}$Univ. Paris-Sud, Univ. Paris-Saclay 91405 Orsay, France \\ and Laboratoire Mat\'eriaux et Ph\'enom\`enes Quantiques, UMR 7162 CNRS, F-75205 Paris, France.}

\begin{abstract}
The connection between the Leggett-Garg inequality and optimal scenarios from the point of view of quantum metrology is investigated for perfect and noisy general dichotomic measurements. In this context, we show that the Fisher information can be expressed in terms of quantum temporal correlations. This connection allows us to associate scenarios with relatively high Fisher information to scenarios in which the Leggett-Garg inequality is violated. We thus demonstrate a qualitative and, to some extent, quantitative link between measurement invasiveness and metrological performance. Finally, we illustrate our results by using a specific model for spin systems.
\end{abstract}
\vskip2pc

\maketitle

\section{Introduction}
The term ``macroscopic" has always been intuitively associated with classical physics.  {\it Macroscopic objects}, for instance, are  the ones observed in our everyday life scale, and are expected to behave according to the laws of classical physics. It is known that classical physics fails to provide a description of phenomena at the microscopic level, which demand the application of quantum mechanical principles, such as the superposition principle and entanglement. Therefore, one is naturally led to the question of whether such quantum mechanical principles could also be observed at the macroscopic scale. This fundamental question concerning the validity of extrapolating quantum mechanics to the macroscopic world \cite{Leggett3} was already pictured in 1935 in the Schr\"odinger's cat {\it Gedanken} experiment \cite{Schrodinger}, where superpositions of states (``dead" and ``alive") of a macroscopic object (the cat) are at stake.

Aiming to propose a test  capable of experimentally ruling out the classical perspective of how macroscopic systems are expected to behave, Leggett and Garg proposed the Leggett-Garg inequality (LGI) \cite{LG, Emary}. The authors considered measurements of two-valued quantities $Q(t_i)$ in macroscopic systems at four different times $\{t_{1},{\cdots},t_{4}\}$. From measurement outcomes, one can compute the correlations $C_{kl} \equiv \langle Q(t_k)Q(t_l)\rangle$. The LGI can be expressed as
\begin{equation}\label{eq1}
-2 \le K_{LG} \equiv C_{12}+C_{23}+C_{34}-C_{14} \le 2,
\end{equation}
and it holds under the following assumptions:
\begin{itemize}
\item[(i)] \textit{macroscopic realism:} a macroscopic system with two or more macroscopically distinct states available to it will at all times be in one of those states; and
\item[(ii)] \textit{noninvasive measurability:} it is possible, in principle, to determine the state of the system with arbitrarily small perturbation to its subsequent dynamics.
\end{itemize}
Therefore, according to Leggett and Garg, the violation of (\ref{eq1}) witnesses the ``{\it nonclassicality}" of the system considered, in line with the definition of classicality provided by (i) and (ii) above.

The LGI and the meaning of its violation have been the subject of recent debates in the literature \cite{Maroney, Clemente, Clemente2, Haliwell, Moreira}. In Ref. \cite{Maroney} it is shown that  only the assumption of noninvasive measurability is tested by the LGI in a model independent way. Given this, in Ref. \cite{Moreira}, some of the authors introduced an operational model relating the LGI violation with a parameter called the {\it measurability} of physical systems. The results were illustrated using perfect and noisy parity measurements performed in  spin-$J$ systems. According to our model,  the more the system is ``measurable,'' i.e., the more one is able to faithfully distinguish between its different possible outcomes, the more the LGI is violated.

Maximum measurability corresponds to projective measurements. As measurability decreases and the measurements become weaker, LGI violation progressively diminishes, eventually vanishing at some point. Therefore, measurability is clearly associated with the invasiveness of measurements, which in turn can depend on e.g. measurement errors or on a dimension-dependent coarse graining \cite{Kofler}. According to this model, the violation of the LGI does not intrinsically depend on the system's size, a notion that lacks itself of precise definition whenever quantum systems are concerned \cite{Yadin, Frowis2, Kwon}. Recently, remarkable experimental achievements as well as experimental proposals regarding LGI violation for systems which can be reasonably considered macroscopic were presented in Refs. \cite{Knee, Formaggio, Budroni}.

Seemingly unconnected, the field of quantum metrology has recently attracted considerable attention \cite{Giovannetti, Paris, Escher, Escher1, Frowis, Mehboudi,Braun,Demkowicz-Dobrzanski}. The use of some quantum mechanical states as probes for the sake of estimating a parameter $\theta$ has been shown to lead to a better scaling, with the dimension of the state, of the precision in the parameter's estimation than using classical resources only. For noisy systems, it was shown that this scaling actually depends on the system's size, the noise parameter, and the noise model \cite{Escher}. Ultimately, for a fixed dimension of the probe state, the precision of the estimation of $\theta$ usually decreases as the noise parameter increases, unless one resorts to appropriate control or error-correcting methods \cite{Braun}.

In light of these elements, it thus seems natural to investigate connections between the LGI violation and quantum metrology. In this work, we do so by identifying each step of the LGI test with the steps of a metrological scenario. Assuming unbiased measurements (implying that the average of the estimated value over all experimental results coincides with the true value of the parameter), we first introduce the definition of \textit{classical} Fisher information (called, from now on, Fisher information), which bounds the standard deviation $ \Delta\theta $ of the estimate of $\theta$ as $\Delta \theta \ge 1/\sqrt{\nu F(\theta)}$ \cite{Fisher, Cramer, Rao}, where $\nu$ is the number of realizations of the experiment. For a given measurement, $ F $ is given by
\begin{equation}\label{eq2}
F(\theta)=\sum_l P_l(\theta)\left[\frac{\partial \ln P_l(\theta)}{\partial \theta}\right]^2,
\end{equation}
where the $P_l(\theta)$ are the probabilities of obtaining each one of the different outcomes $l$ and thus, $\sum_l P_l(\theta)=1$. The generalization of the Fisher information to quantum mechanics is done by writing $P_l (\theta)={\rm Tr}[\rho(\theta)E_l]$, where ${E_l}$ is a \textit{positive operator valued measure} (POVM). By maximizing $F(\theta)$ over all quantum measurements, one obtains the quantum Fisher information (QFI) $\mathcal{F}_Q$ \cite{Helstrom, Holevo, Braunstein, Braunstein2}, associated with the minimum lower bound for $\Delta\theta$, and saturated when $\nu\rightarrow\infty$. Hence, the QFI corresponds to the Fisher information associated with the optimal measurement, i.e., the one which gives the best estimation for $\theta$.

\section{LGI and metrological protocols}
In this section we compare the LGI test scenario to a parameter estimation protocol and establish some general results.
In a LGI test, a maximally mixed initial state $\hat \rho_0= \mathbb{I}/d$ is prepared, where $d$ is the dimension of the underlying Hilbert space.
The dichotomic observable measured in the LGI is denoted by $\hat A$, and the unitary time evolution, generated by the Hamiltonian $\hat H$, $\hat U (t_i) = e^{-iHt_i}$ (in what follows $ \hbar = 1 $). From now on, we suppose that, by rescaling the energies, the times $t_i$ are dimensionless.
Using these definitions, we have that the two-time correlation appearing in the LGI~\eqref{eq1} can be written as $C_{ij}={\rm Tr}[  \hat A \hat U(t_j-t_i) \hat A \hat U(t_i) \hat \rho_0 \hat U_i^{\dagger}(t_i) \hat U_j^{\dagger}(t_i-t_j) ] = C(\theta_{ij})$, where $\theta_{ij} \equiv t_i-t_j$.

 From now on we take all the time intervals $t_2-t_1=t_3-t_2=t_4-t_3\equiv \theta$ to be equal. While our results can be extended to the case of different time intervals, this choice is particularly convenient since it will allow us to express the Leggett-Garg parameter $K_{LG}$ as a simple function of the parameter $\theta$, which will be the object of interest in our metrological investigation (as detailed in the following). Under this specification, we can then write the correlation function $C(\theta)$ as
\begin{equation}
C(\theta) = \frac{1}{d}{\rm Tr}[  \hat A \hat U(\theta) \hat A \hat U^{\dagger}(\theta) ].
 \end{equation}

Generally, to calculate the LGI~\eqref{eq1}, one must perform four independent experiments in order to measure each one of the correlations $C_{12}$, $C_{23}$, $C_{34}$ and  $C_{14}$. However, since we assumed all equal time intervals, the correlation functions $C_{ij}$ are  {\it stationary} \cite{Emary}, that is, they depend only on the time difference $\theta$. The LGI can in fact be rewritten as
\begin{equation}\label{LGIbis}
\vert K_{LG}(\theta) \vert = |3C(\theta)-C(3\theta)|\leq 2,
\end{equation}
meaning that it suffices to determine only two terms: $C(\theta)$ and $C(3\theta)$. Therefore, only two independent experiments, in which one performs two subsequent measurements, are required in this case.

Since we start from the maximally mixed state, the system will remain unchanged before the first measurement. After it, however, the system's state will be one of the two possible outcomes resulting from the measurement of $\hat A$, i.e. either $ \hat{\rho}_+ $ or $ \hat{\rho}_- $.  We shall thus refer to the first measurement as the {\it preparation procedure}.

We now introduce a metrological scenario which can be related to the LGI protocol described above. We consider the problem of estimating the unknown parameter $\theta$ through a measurement of the same dichotomic observable $\hat{A}$. This second measurement can be either projective or noisy, and may be generally described by a two-valued POVM. Also because we consider a maximally mixed initial state, the result of the first measurement is symmetric. That is, each one of its possible outcomes can be obtained with equal probabilities $1/2$. At the time the second measurement is performed, the evolved system's state is $\hat\rho_\pm(\theta)=\hat U(\theta)\hat \rho_{\pm}\hat U^\dagger(\theta)$.

The precision of the estimation can be characterized by the Fisher information $F(\theta)$ as given by Eq.~\eqref{eq2}, in which $P_l(\theta)$, with $l=\pm 1$, is the probability for measuring $\hat\rho_\pm(\theta)$. As shown in the Appendix \ref{appendixA}, both $ \hat{\rho}_+ $ and $ \hat{\rho}_- $ yield identical $ F(\theta) $. The latter may be written in terms of the correlation function $ C(\theta) $ as
\begin{equation}\label{FisherC}
F(\theta)=\frac{1}{1-C(\theta)^2} \left [  \frac{ \partial C(\theta)} {\partial \theta}\right ]^2.
\end{equation}
We note that Eq.~\eqref{FisherC} does not depend on the specific assumption of equal time intervals  $\theta$, and can be straightforwardly generalized to construct the Fisher information matrix $F_{ij}$ as a function of the correlation functions $C_{ij}$ for the multi-parameter estimation of all the time separations $\theta_{ij}$. However, as anticipated we will focus on the particularly instructive case of a single parameter $\theta$. Equation~\eqref{FisherC} turns out to have several remarkable properties and serves as a guideline to establish a connection with the LGI.

First, we note that the extrema of $C(\theta)$ are also the extrema of $F(\theta)$. {\it A priori}, the former are not all extrema of the latter, but let us focus on their common extrema, which we will label by $\theta_e$. It is straightforward to show that $\theta_e$ corresponds to a maximum of $F$ if and only if $C(\theta_e)^2=1$.
The value $C=\pm 1$ can only be obtained for an ideal projective measurement. In particular, if we have such an extremum of $C$ at $\theta=0$,
and if $C(\theta)$ is a periodic function  with the period denoted by $T$ (which is the case if the Bohr frequencies of $H$ are commensurate), then $\theta = nT$ ($n\in \mathbb N$) will also correspond to extrema. In this last case, as $C(\theta)$ is an even function of $ \theta $, it can be shown that  $\theta = nT/2$ is also an extremum of $C(\theta)$. The global extremum corresponds to the value $C(nT/2)^2=1$ only for ideal projective measurements.

On the other hand, if an extremum of $C$ is such that  $C(\theta_e)^2 \neq 1$, then $F(\theta_e)=0$. Therefore, we find a very peculiar situation, in which the estimation of $\theta_e$ can be optimal when the measurement is ideally projective but all information about $\theta_e$ is lost when an infinitesimal amount of noise is added to the measurement (i.e., if $C(\theta_e) = 1 - \epsilon$, then $F(\theta_e) = 0$ for arbitrarily small $\epsilon$). In other words, the maximum of $F$ which is also an extremum of $C$ is not robust against noisy measurements for parameter estimation.

\section{Parity measurement on a spin system}
Equation (\ref{FisherC}) is quite general, and is based only on the fact that dichotomic measurements are performed in order to estimate the parameter $\theta$. In the following, we will consider a specific example which illustrates its consequences.

We study the case of parity measurements performed in a spin-$J$ system. Parity has been shown to  be useful in quantum optical metrology \cite{Chiruvelli, Gerry}, and has also been used in \cite{Moreira} as part of a model where the LGI violation is controlled through a parameter determining the invasiveness of a POVM.
Let us briefly recall the main properties of this model.
We  consider a spin operator $\hat{J}$, with spatial components $\hat{J}_{\upsilon}$, $\upsilon=x, y, z$. The $\hat J_z$ eigenstates are denoted as $\left|m\right>$ , $-j \leq m \leq j$, where $j(j+1)$ ($j\in \mathbb N$) are the eigenvalues of $\hat{\bf J}^2$.  The dynamics of the system is governed by the following Hamiltonian:
\begin{equation}\label{eq:eq3}
\hat{H}=\Omega \hat{\bf{J}}^2+\omega\hat{J}_x,
\end{equation}
where $\Omega$ and $\omega$ are constants  with the dimension of frequency.
In our setting, the initial state is
\begin{equation}\label{eq:eq4}
\rho_0=\frac{1}{2j+1}\sum_{m=-j}^j\left|m\right>\left<m\right|,
\end{equation}
so that LGI violations can only arise from the measurements and system's dynamics.
We consider the two-valued POVM introduced in Ref. \cite{Moreira}
\begin{equation}\label{povm}
\hat{E}_{\pm}=\hat{M}_{\pm}^\dagger\hat{M}_{\pm}=\frac{1}{2}(\mathbb{I}\pm\hat{A}),
\end{equation}
where the dichotomic observable $ \hat{A} $ takes the form
\begin{equation}\label{eq:eq6}
\hat{A}\equiv  \sum_{\mu}\sum_{m \in \Delta m_{\mu}}(-1)^{(j-m)}f_{{\mu}}(m, \sigma)\left|m\right>\left<m\right|.
\end{equation}
The functions $f_{{\mu}}(m, \sigma)=e^{\frac{-(m-\mu)^2}{2\sigma^2}}$ and $\Delta m_\mu$ are disjoint sets containing equally sized intervals of $m$. The parameter $\sigma$ can be interpreted as being associated with the unfaithfulness of the measurement: for finite $\sigma$ and $m \neq \mu$, the particle is detected, but the value of $m$ cannot be perfectly determined. Hence, $\sigma \rightarrow \infty$ implies performing projective measurements, with perfect determination of the system's parity as, for this case, $\hat A=\hat \Pi_z=\sum_m(-1)^{j-m} \left|m\right>\left<m\right|$.  Finally, the parameter $\Delta m_{\mu}$ determines the number $N$, among all the possible values of $m$ that the measurement apparatus can faithfully detect, and therefore is called the {\it resolution}.

We now study numerically the example of a spin $5/2$ in order to illustrate our results. In this example, we introduce a parameter $b$ that is directly associated with the measurability of the system or, alternatively, with the invasiveness of a measurement and the width $\sigma$ of the function $f_{{\mu}}(m, \sigma)=e^{\frac{-(m-\mu)^2}{2\sigma^2}}$. By defining  $b\equiv e^{-1/2\sigma^2}$, we have that  $\sigma\rightarrow \infty$ corresponds to $b\rightarrow 1$, and $\sigma\rightarrow 0$ to $b\rightarrow 0$. In Eq.~\eqref{eq:eq6}, we have considered only two possible values of $\mu$, $\mu_{\pm}=\pm 5/2$, and the two corresponding intervals are $\Delta m_{\mu_{-}}=[-5/2, 0)$ and $ \Delta m_{\mu_{+}} = (0,5/2] $.

We move on to the computation of $ F(\theta) $ and $ \mathcal{F}_Q $. As mentioned before, the first parity measurement of the LGI is identified as the state preparation in the quantum metrology protocol. The resulting state after this first measurement is given by one of the two states:
\begin{equation}\label{eq7}
\hat{\rho}_{\pm}(t_k)=\frac{(\hat{E}_{\pm})^{\frac{1}{2}}\hat{\rho}_0(\hat{E}_{\pm})^{\frac{1}{2}}}{p_{\pm}},
\end{equation}
where $p_{\pm} = {\rm Tr}(\hat{E}_\pm\hat{\rho}_0)$.

Without loss of generality, we will work with $\hat{\rho}_+$ ($\hat{\rho}_-$ gives the same results). According to Eq.~\eqref{eq:eq3}, the evolved state $\hat{\rho}_+(\theta)$, before the realization of the second parity measurement can be written as
\begin{equation}\label{eq8}
\hat{\rho}_+(\theta)=\hat{U}(\theta)\hat{\rho}_+\hat{U}^\dagger(\theta)=e^{-i\theta\hat{J}_x}\hat{\rho}_+ e^{i\theta\hat{J}_x}.
\end{equation}
Combining Eqs.~\eqref{eq2} and \eqref{povm} we evaluated the Fisher information $ F(\theta) $ and the QFI $\mathcal{F}_Q$ \cite{Liu}. In the following, we split the analysis into two cases: projective and noisy parity measurements.

\begin{figure}[h]
\includegraphics[width=8.5cm]{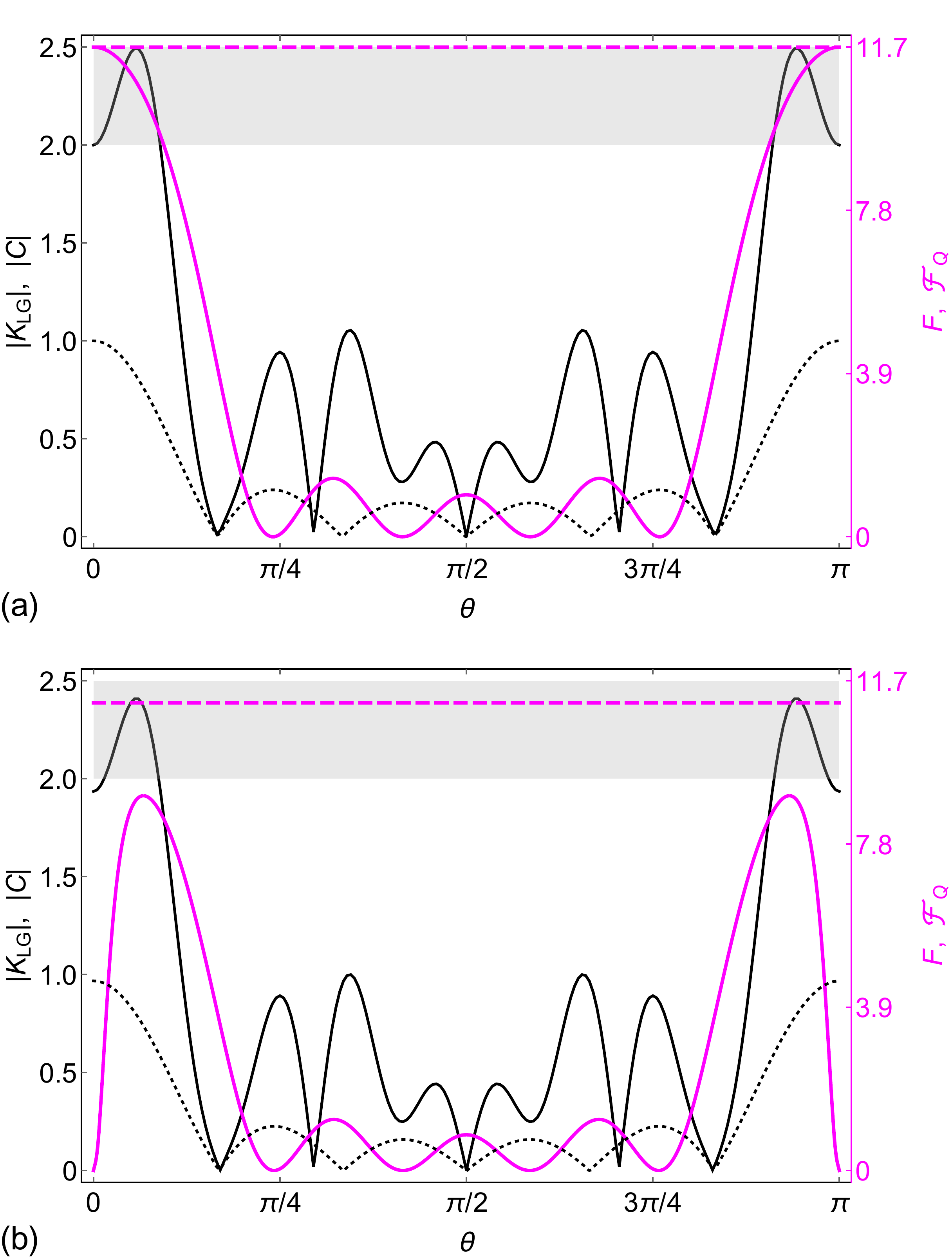}
\caption{(color online) Plots of the Fisher information $F$ (solid magenta line, scaled to the right vertical axis), the quantum Fisher information $\mathcal{F}_Q$ (dashed magenta line, scaled to the right vertical axis), the absolute value of the two-time correlation $C$ (dotted black line, scaled to the left vertical axis) and absolute value of $K_{LG}$ (solid black line, scaled to the left vertical axis), as a function of $\theta$, for (a) $b=1$ and (b) $b=0.99$. The LGI violation region (relative to the left vertical axis) is shaded in light gray. All the plotted quantities are dimensionless. }\label{fig:projective}
\end{figure}

\subsection{Projective parity measurements ($ b = 1 $)}

The results for $b=1$, i.e. for noise-free parity measurements, are shown in Fig.~\ref{fig:projective}(a). The Fisher information $F(\theta)$ and the quantum Fisher information $\mathcal{F}_Q$ are both plotted in this figure. We see that they coincide for $\theta=n\pi$, showing that the measurement scenario is optimal at this point.
We also note that these maxima of  $F(1,\theta)$ are also extrema of $C(\theta)$, and, as expected, the correlation function reaches its optimal value $C=\pm1$ at these points.
We then compare these results to $K_{LG}$ defined in Eq.~\eqref{eq1} as a function of $\theta$. The point of maximal correlation cannot be a point of LGI violation, and this is well illustrated in Fig.~\ref{fig:projective}. Therefore, invasiveness cannot be witnessed for $\theta=n\pi$. As we can see from Fig.~\ref{fig:projective}(a), the region around $\theta=n\pi$ corresponds to relatively high Fisher information, and maximum LGI violation also occurs in this region. Therefore, the most favorable metrological scenario occurs in the same region where invasiveness is witnessed through LGI violation. Nevertheless, the maximum of the Fisher information does not coincide with the maximum violation of the LGI.

\subsection{Noisy parity measurements ($ b < 1 $)}

We now examine the cases corresponding to limited precision, which corresponds to measurability $b < 1$. As $b$ decreases and the measurements become noisier, both LGI violation and the optimality of the metrological scenario are progressively degraded.
In Fig.~\ref{fig:projective}(b), we have plotted $F(\theta)$, $C(\theta)$ and $K_{LG}(\theta)$ for $b=0.99$. We now observe that the Fisher information is zero at $\theta=n\pi$. Recall that this drastic transition follows from Eq.~\eqref{FisherC}: as discussed above, the ``collapse" of the Fisher information under the addition of noise occurs because  $\theta=n\pi$ correspond to common extrema of $C$ and $F$. This observation further suggests that the LGI violation at the maximum Fisher information is a hallmark of the robustness of the latter against noise.

In this way, we have obtained, in the framework of this specific model, a connection between the points where invasiveness is witnessed and those corresponding to favourable and noise-robust metrological scenarios.

\begin{figure}[h]
\ \includegraphics[width=7.65cm]{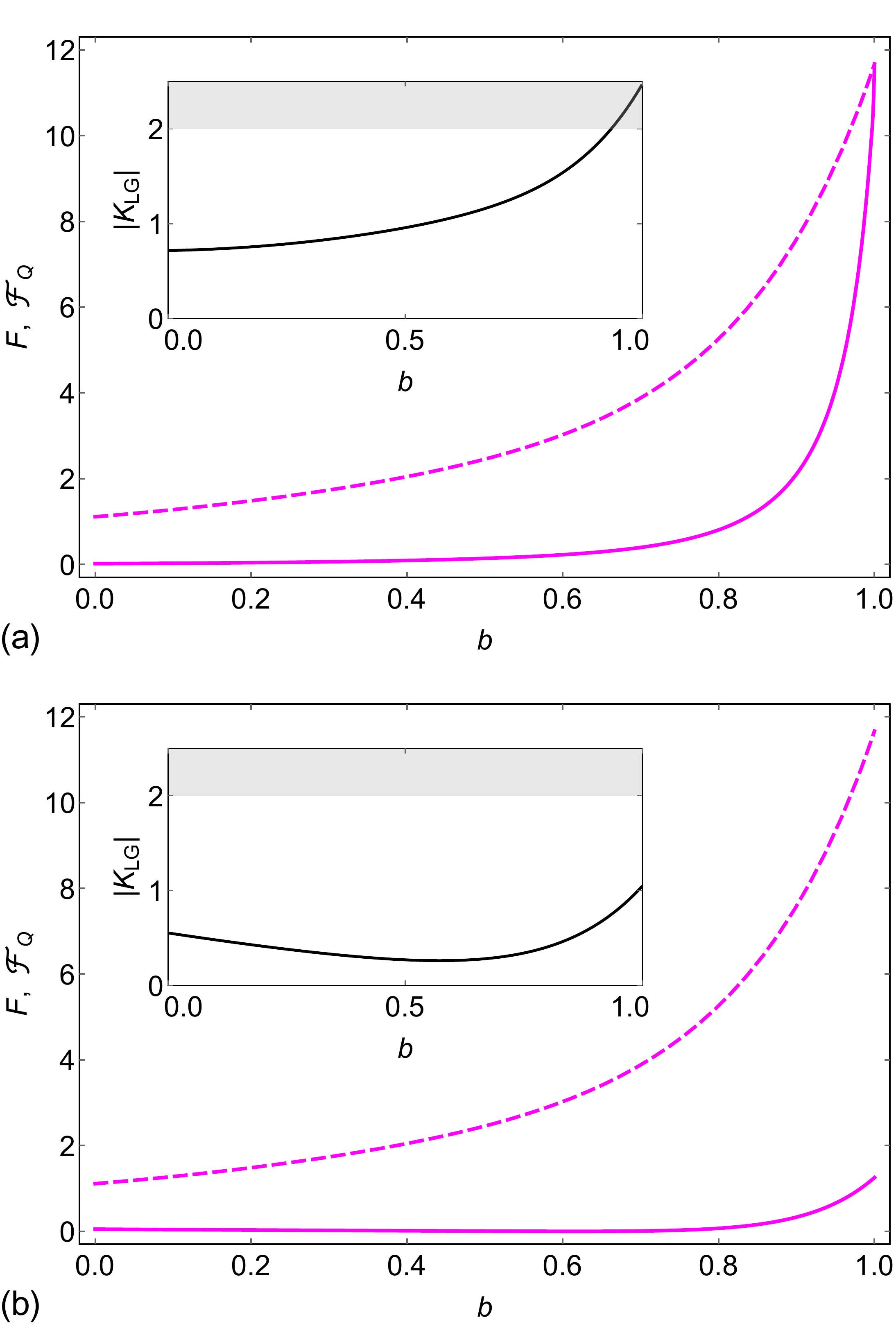}\hfill \quad
\caption{(color online) Plots of the Fisher information $F$ (solid magenta line), the quantum Fisher information $\mathcal{F}_Q$ (dashed magenta line), and the absolute value of the Leggett-Garg parameter $|K_{LG}|$ (inset, solid black line), as a function of $b$ for (a) $\theta/\pi=0.95$ and (b)  $\theta/\pi=0.34$. In the insets, the LGI violation region is shaded in light gray; note that the LGI is violated in the interval $0.94 \le b \le 1$ for case (a), while it is never violated for case (b). All the plotted quantities are dimensionless.}\label{fig:fig1}
\end{figure}

\section{Discussion}
Our model sheds light on the relationship between the quantum Fisher information and quantum invasiveness. Some physical insight about this connection has already been given in Ref. \cite{Frowis} where, by taking into account a ``no-signaling in time condition'' \cite{Kofler2}, the authors argued that quantum states with large $ \mathcal{F}_Q $ are necessary for LGI violation with large measurement uncertainties.

In order to further investigate this point, we plotted $F$, $\vert K_{LG} \vert$ and $\mathcal{F}_Q$ as a function of $b$ for fixed values of $\theta$ in Fig.~\ref{fig:fig1}.
Specifically, in Fig.~\ref{fig:fig1}(a), we have fixed $\theta/\pi=0.95$, a value that allows the violation of the LGI for $b>0.94$. We see that both $\mathcal{F}_Q$ and
$F$ increase monotonically as $b$ increases and $F$ approaches its optimal value, $\mathcal{F}_Q$, in the region where LGI is violated. On the other hand, in Fig.~\ref{fig:fig1}(b), we take $\theta/\pi=0.34$ so that no violation of the LGI can occur. Note that $\mathcal{F}_Q$ remains the same as a function of $b$, as $\mathcal{F}_Q$  does not depend on $\theta$. It is thus clear that large QFI is not a sufficient condition for violation of the LGI.

\begin{figure}[h]
\centering 
\includegraphics[width=8.5cm]{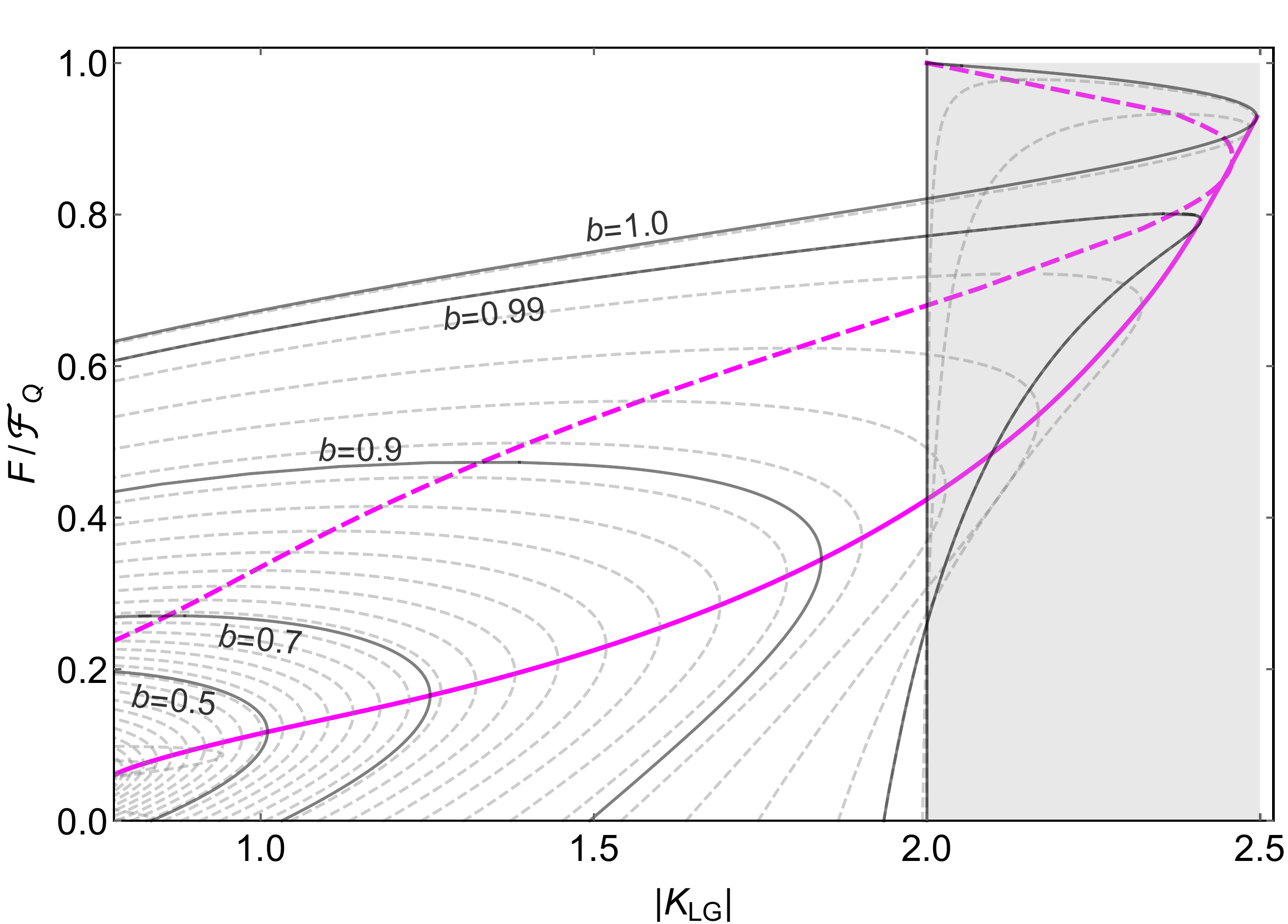}
\caption{(color online) Normalized Fisher information, $F/\mathcal{F}_Q$, versus the absolute value of the Leggett-Garg parameter $|K_{LG}|$. The dashed gray lines are contours at fixed $b$ for all $\theta \in [0, \pi/2]$. Specific contours at $b=0.5, 0.7, 0.9, 0.99, 1.0$ are highlighted as solid dark gray lines.  The solid magenta line connects the points at the optimal $\theta$ maximizing the LGI violation, while the dashed magenta line connects the points at  the $\theta$ maximizing instead the Fisher information. The LGI violation region is shaded in light gray. All the plotted quantities are dimensionless.}\label{LGFisher}
\end{figure}

Note as well that, in Fig.~\ref{fig:fig1}(b), the Fisher information increases monotonically as $b$ increases but it does not reach its optimal value, $\mathcal{F}_Q$. In order to explore in further detail the quantitative relationship between the Fisher information and the LGI violation, we plot in Fig.~\ref{LGFisher} the normalized Fisher information $F/\mathcal{F}_Q$ versus the absolute value of the Leggett-Garg parameter $|K_{LG}|$ (see caption for details).
We see that both the maximum (normalized) Fisher information and the maximum of $|K_{LG}|$ monotonically increase with $b$. Furthermore, it is interesting to note that $ F/\mathcal{F}_Q $ at the point of maximal violation of the LGI (solid magenta line) is a monotonically increasing function of the violation itself. We can also see that, even though the maximization of $F$ and $|K_{LG}|$ are generally incompatible, violation of the LGI is necessary to access the nearly optimal regime of $F/{\cal F}_Q$ above $\approx 0.82$. Finally, we see that, when the LGI is violated, there is a lower bound for the Fisher information, given by $F/{\cal F}_Q \gtrsim 0.27$, thus LGI violation guarantees a non-trivial minimum metrological precision. Conversely, when the LGI is not violated, the Fisher information can be arbitrarily small and vanish for specific parameter settings.

\section{Conclusion}
We have established a connection between temporal correlations, involved in Leggett-Garg inequality tests, and the Fisher information associated with a specific metrological scenario. In particular, guided by the general expression of the Fisher information in terms of two-time correlation functions, we established that the precision of the estimation is very fragile against noise unless accompanied by LGI violation. In addition, and looking at a specific example, we showed that a large quantum Fisher information is not sufficient for violating the LGI. We also illustrated how a violation of the LGI may set a non-trivial lower bound to the precision of parameter estimation while, on the other hand, large LGI violations may enable nearly optimal parameter estimation. The ultimate precision limit in which Fisher information and quantum Fisher information coincide may only be achieved in the presence of a violation of the LGI. Generalizations of such intriguing connections between measurement invasiveness and sensitivity beyond specific models certainly deserve further investigation.

\section*{Acknowlegments}
The authors acknowledge L.~Davidovich for insightful discussions. S.~V.~M.~acknowledges financial support from the Brazilian agency CAPES. L.~A.~C.~and G.~A.~acknowledge financial support from the European Research Council under the StG GQCOP (Grant No.~637352) and the Royal Society under the International Exchanges Programme (Grant No.~IE150570).

\appendix

\section{}\label{appendixA}

In the following, we will show that Eq.~\eqref{FisherC} can be derived by considering either $\rho_+(\theta)$ or $\rho_-(\theta)$ as the preparation state. Recall that we can express the two-valued POVMs as
\begin{equation}
\hat{E}_{\pm}=\hat{M}_{\pm}^\dagger\hat{M}_{\pm}=\frac{1}{2}(1\pm\hat{A}).
\end{equation}

In this way, if  $\rho_+(\theta)$ is considered, the probabilities of obtaining the outcomes $\pm$ at the time at which the second measurement is performed can be written as
\begin{equation}\label{ProbMetrology}
P_{\pm}(\theta) = {\rm Tr}(\hat{E}_{\pm}\rho_+(\theta)) = \frac{1}{2} \pm \frac{1}{2} C(\theta),
\end{equation}
and, if one considers the preparation $\rho_-(\theta)$, we have
\begin{equation}\label{ProbMetrology2}
P_{\pm}(\theta) = {\rm Tr}(\hat{E}_{\pm}\rho_-(\theta)) = \frac{1}{2} \mp \frac{1}{2} C(\theta).
\end{equation}

One obtains Eq.~\eqref{FisherC} by considering either the probability distribution \eqref{ProbMetrology} or \eqref{ProbMetrology2} in Eq.~\eqref{eq2}.


\begin{thebibliography}{99}

\bibitem{Leggett3} A. J. Leggett, J. Phys. Condens. Matt. {\bf 14}, R415 (2002).
\bibitem{Schrodinger} E. Schrodinger, Naturwissenschaften, {\bf 23}, 807 (1935).
\bibitem{LG} A. J. Leggett  and A. Garg, Phys. Rev. Lett. {\bf 54}, 857 (1985).
\bibitem{Emary} C. Emary, N. Lambert, F. Nori, Rep. Prog. Phys. {\bf 77}, 016001 (2014).
\bibitem{Maroney} O. J. E. Maroney and C. G. Timpson, arXiv: 1412.6139v1 (2014).
\bibitem{Clemente} L. Clemente and J. Kofler, Phys. Rev. A {\bf 91}, 062103 (2015).
\bibitem{Clemente2} L. Clemente and J. Kofler, Phys. Rev. Lett. {\bf 116}, 150401 (2016).
\bibitem{Haliwell} J. J. Haliwell, Phys. Rev. A {\bf 93}, 022123 (2016).
\bibitem{Moreira} S.V. Moreira, A. Keller, T. Coudreau, and P. Milman, Phys. Rev. A {\bf 92}, 062132 (2015).
\bibitem{Kofler} J. Kofler and C. Brukner, Phys. Rev. Lett. {\bf 99}, 180403 (2007).
 \bibitem{Yadin} B. Yadin and V. Vedral, Phys. Rev. A {\bf 93}, 022122 (2016).
\bibitem{Frowis2} F. Frowis and W. Dur, New J. Phys. {\bf 14}, 093039 (2012).
\bibitem{Kwon} H. Kwon, C-.Y. Park, C. Tan and H. Jeong, New J. Phys. {\bf 19}, 043024 (2017).
\bibitem{Knee} G. C. Knee, K. Kakuyanagi, M-.C. Yeh, Y. Matsuzaki, H. Toida, H. Yamaguchi, A. J. Leggett and W. J. Munro, Nat. Commun. {\bf 7}, 13253 (2016).
\bibitem{Formaggio} J. A. Formaggio, D. I. Kaiser, M. M. Murskyj, T. E. Weiss, Phys. Rev. Lett. {\bf 117}, 050402 (2016).
\bibitem{Budroni} C. Budroni, G. Vitagliano, G. Colangelo, R.J. Sewell, O. G\"uhne, G. T\'oth, Phys. Rev. Lett. {\bf 115}, 200403 (2015).
\bibitem{Giovannetti} V. Giovannetti, S. Lloyd, L. Maccone, Science, {\bf 306}, 1330 (2004).
\bibitem{Escher} B. M. Escher, R. L. de Matos Filho, and L. Davidovich, Nat. Phys. {\bf 7}, 406 (2011).
\bibitem{Escher1} B. M. Escher, R. L. de Matos Filho, and L. Davidovich, Braz. J. Phys. {\bf 41}, 229 (2011).
\bibitem{Paris} M. G. A. Paris, Int. J. Quant. Inf. {\bf 7}, 125 (2009).
\bibitem{Frowis} F. Fr\"owis, P. Sekatski and W. D\"ur, Phys. Rev. Lett. {\bf 116}, 090801 (2016).
\bibitem{Mehboudi} M. Mehboudi, L. A. Correa and A. Sanpera, Phys. Rev. A {\bf 94}, 042121 (2016).
\bibitem{Demkowicz-Dobrzanski} R. Demkowicz-Dobrzanski, J. Kolodynski, M. Guta, Nature Comm. {\bf 3}, 1063 (2012).
\bibitem{Braun} D. Braun, G. Adesso, F. Benatti, R. Floreanini, U. Marzolino, M. W. Mitchell, S. Pirandola, arXiv: 1701.05152.
\bibitem{Fisher} R. A. Fisher, Messenger of Mathematics {\bf 41}, 155 (1912).
\bibitem{Cramer} H. Cram\'er, {it Mathematical Methods of Statistics} (Princeton University Press, Princeton, 1946).
\bibitem{Rao} C. R. Rao, {\it Linear Statistical Inference and its Applications}, 2nd ed. (Wiley, New York, 1973).
\bibitem{Holevo} A. S. Holevo, {\it Probabilistic and statistical aspects of quantum theory}, (North-Holand, Amsterdam, 1982).
\bibitem{Helstrom} C. W. Helstrom, {\it Quantum detection and estimation theory}, (Academic Press, New York, 1976).
\bibitem{Braunstein} S. L. Braunstein, C. M. Caves and G. J. Milburn,  Ann. Phys. {\bf 247},135 (1996).
\bibitem{Braunstein2} S. L. Braunstein and C. M. Caves, Phys. Rev. Lett. {\bf 72}, 3439 (1994).
\bibitem{Chiruvelli} A. Chiruvelli and H. Lee, J. Mod. Opt. {\bf 58}, 945 (2011).
\bibitem{Gerry} C. C. Gerry and J, Mimih, Contemp. Phys. {\bf 51}, 497 (2010).
\bibitem{Liu} L. Jing, J. Xiao-Xing, Z. Wei and W. Xiao-Guang, Comm. Theor. Phys., {\bf 61}, 45 (2014).
\bibitem{Kofler2} J. Kofler and C. Brukner, Phys. Rev. A {\bf 87}, 052115 (2013).

\
\end{thebibliography}
\end{document}